\documentclass[aps,prd,preprintnumbers,superscriptaddress,nofootinbib,notitlepage,twocolumn]{revtex4-2}
\usepackage[pdftex]{graphicx}
\usepackage{bm,latexsym,amsmath,amssymb,amsfonts,mathrsfs}
\usepackage{color}
\allowdisplaybreaks[1]
\usepackage[pdftex,colorlinks=true,linkcolor=blue,citecolor=cyan]{hyperref}
\newcommand*{\D}{{\rm d}}
\newcommand*{\mpl}{M_{\rm Pl}}
\newcommand*{\hinf}{H_{*}}
\newcommand*{\ga}{g_{A}}
\newcommand*{\psim}{\psi_{\rm min}}
\newcommand\figref[1]{Fig.~\ref{#1}}
\newcommand\tabref[1]{Table~\ref{#1}}

\begin{document}
\title{Dynamics of inflation with mutually orthogonal vector fields in a closed universe}
\author{Tomoaki~Murata}
\email[Email: ]{tmurata@rikkyo.ac.jp}
\affiliation{Department of Physics, Rikkyo University, Toshima, Tokyo 171-8501, Japan
}
\author{Tsutomu~Kobayashi}
\email[Email: ]{tsutomu@rikkyo.ac.jp}
\affiliation{Department of Physics, Rikkyo University, Toshima, Tokyo 171-8501, Japan
}
\begin{abstract}
We study the dynamics of a homogeneous, isotropic, and positively curved universe in the presence of a SU(2) gauge field or a triplet of mutually orthogonal vector fields.
In the SU(2) case we use the previously known ansatz for the
gauge-field configuration, but the case without
non-abelian symmetries
is more nontrivial and we develop a new ansatz.
We in particular consider axion-SU(2) inflation and inflation with
vector fields having U(1)$\times$U(1)$\times$U(1) symmetry,
and analyze their dynamics in detail numerically.
Novel effects of the spatial curvature come into play through
vector fields, which causes unconventional pre-inflationary dynamics.
It is found that the closed universe with vector fields is slightly more stable
against collapse than that filled solely with an inflaton field.
\end{abstract}
\preprint{RUP-21-12}
\maketitle
\section{Introduction}

Inflation~\cite{Starobinsky:1980te,Guth_1981,Sato:1980yn} is
an accelerated expansion that occurred in the early Universe.
It can explain homogeneity, isotropy, and spatial flatness of the Universe
in a natural way.
Furthermore, it can give rise to primordial fluctuations that are
consistent with observations of
the CMB and large-scale structure of the Universe.
Inflation was thus introduced to resolve the problems on the
initial conditions in the standard big-bang model.
This does not, however, necessarily mean that
the Universe is homogeneous, isotropic, and spatially flat
at around the beginning of inflation.
Therefore, it is important to assess to what extent
inflation is likely to occur and thus to erase inhomogeneity,
anisotropy, and spatial curvature
starting from generic initial conditions.

In spite of its importance, not so many papers
(as compared to a huge number of papers on inflation)
have been devoted to the problem of the initial conditions for inflation.
For example, Ref.~\cite{Wald:1983ky} showed that
homogeneous and anisotropic Bianchi models except Bianchi type-IX universes
always evolve toward an isotropic attractor
in the presence of a positive cosmological constant.
The exceptional case corresponds to the universe with
positive spatial curvature; the universe would collapse if
the spatial curvature is as large as the cosmological constant.
As shown in Ref.~\cite{Belinsky:1987gy}, the positive spatial curvature
reduces (but not significantly\footnote{The question of how much is
significant is a difficult one because in principle
the phase space of initial conditions would be infinite
and a careful consideration is necessary.
It is still interesting to consider whether certain fields
make the process more robust, but defining a fraction is
difficult when the denominator is infinite.
In this sense, any change that does not render the whole phase
space stable could be considered insignificant.})
the fraction of the initial conditions that lead to successful inflation
in the case of a massive scalar inflaton field (see also Ref.~\cite{Belinsky:1985zd}).
The effect of the initial inhomogeneity on the onset
of inflation has been addressed in Refs.~\cite{Albrecht:1985yf, KurkiSuonio:1987pq, Goldwirth:1989vz, Goldwirth:1989pr, KurkiSuonio:1993fg, East:2015ggf, Brandenberger:2016uzh, Clough:2016ymm, Marsh:2018fsu, Aurrekoetxea:2019fhr}.
More recently, the problem of the initial inhomogeneity
was investigated in the context of multi-field inflation,
with somewhat nontrivial results~\cite{Easther:2014zga}.

While most of the inflationary models are based on one or more scalar fields,
models with a triplet of vector fields have been proposed recently.
Vector fields are apparently incompatible with
isotropic cosmology, but by assuming
a triplet of mutually orthogonal vector fields
one can achieve an isotropic configuration.
An earlier model of inflation driven by such vector fields
nonminimally coupled to gravity
is given in Ref.~\cite{Golovnev:2008cf}.
An interesting example in this class of models
motivated by particle physics is
chromo-natural (or axion-SU(2)) inflation~\cite{Adshead:2012kp}
(see Ref.~\cite{Maleknejad:2012fw} for a review).
Cosmological models with
multiple generalized Proca fields~\cite{Jimenez:2016upj,Allys:2016kbq,Emami_2017,GallegoCadavid:2020dho}
and three copies of U(1) vector fields~\cite{Gorji_2020}
have also been considered in the literature.
See also Refs.~\cite{Armendariz-Picon:2004say,Rodriguez:2017wkg,BeltranJimenez:2018ymu,Alvarez:2019ues,Maleknejad:2011jw,Maleknejad:2011sq,Maleknejad:2011jr,Nieto:2016gnp,Adshead:2016omu,Adshead:2017hnc,Gomez:2019tbj,Guarnizo:2020pkj,Gomez:2020sfz,Bento:1992wy,Maeda:2012eg,Yamamoto:2012sq,Funakoshi:2012ym} for models of similar kinds.

The problem of initial conditions in
the above inflationary models with vector fields
is more subtle and less studied than that in
usual inflation where only a scalar field participates in
the dynamics.
In the context of the axion-SU(2) model,
Bianchi type-I anisotropic cosmology has
been discussed~\cite{Maleknejad:2013npa,Wolfson:2020fqz},
with the conclusion that the initial anisotropies always dilute away
immediately~\cite{Wolfson:2021fya}.
However, the dynamics of the other Bianchi types
of the axion-SU(2) system is still unclear
(see however Ref.~\cite{Maeda:2013daa} for the dynamics of Bianchi universes
in the presence of an SU(2) gauge field coupled to a particular scalar field theory).
The problem of initial conditions in the other vector field models
is also awaited to be explored.


To take a step forward, in this paper, we explore
the spatially curved generalization of homogeneous and
isotropic cosmologies in the axion-SU(2) model and
in (certain variants of) multiple generalized Proca theories,
and study the dynamics of the pre-inflationary Universe.
It is not trivial to introduce a triplet of dynamical vector fields
in such a way that their configuration
is consistent with homogeneity and isotropy in a spatially curved universe.
In fact, the consistent ansatz for the SU(2) gauge field configuration
in a curved universe has long been known~\cite{Hosotani:1984wj,Galtsov:1991un,Galtsov:2008wkj}.
However, the ansatz introduced in Ref.~\cite{Hosotani:1984wj,Galtsov:1991un,Galtsov:2008wkj}
relies on the non-abelian-specific structure,
and hence it cannot be extended straightforwardly
to the case, for example, of the vector field model
with the U(1)$\times$U(1)$\times$U(1) symmetry~\cite{Gorji_2020}.
This point is also addressed in the present paper.

The paper is organized as follows.
In Sec.~\ref{sec:su2curve},
we add the spatial curvature to the axion-SU(2) model,
and then discuss its consequences on the cosmological dynamics
on the basis of the analytic argument and the results of numerical calculations.
In Sec.~\ref{sec:proca}, we consider
spatially curved cosmological models
in the presence of a triplet of (generalized) Proca fields,
with a focus on two particular examples in the literature.
Finally, we draw our conclusions in Sec.~\ref{sec:conclusions},
with a comment on the extension of the present work
to the Bianchi type-IX geometry.

\section{Axion-SU(2) in a Curved Universe}\label{sec:su2curve}

\subsection{Basic equations}

The axion-SU(2) inflation model~\cite{Adshead:2012kp} is described by
the Lagrangian
\begin{align}
{\cal L}_{\rm CN}&=\frac{\mpl^2}{2}R -\frac{1}{4}F_{\mu\nu}^a F_a^{\mu\nu}
-\frac{1}{2}(\partial\chi)^2-V(\chi)
\notag \\ &\quad
-\frac{\lambda}{4f}
\chi\widetilde F_{\mu\nu}^a F^{\mu\nu}_a,
\label{eq:Lag01}
\end{align}
where $\chi$ is the axion field,
\begin{align}
    V(\chi)&=\mu^4\left(1+\cos\frac{\chi}{f}\right),
\end{align}
is its potential, and $F_{\mu\nu}^a$ is the field strength defined as
\begin{align}
F_{\mu\nu}^a=\partial_\mu A_\nu^a-\partial_\nu A_\mu^a+ g_A
\epsilon^a_{\; bc}A_\mu^bA_\nu^c.
\end{align}
The last term in Eq.~\eqref{eq:Lag01} is written more explicitly as
\begin{align}
    \widetilde F_{\mu\nu}^a F^{\mu\nu}_a&=\frac{1}{2}
    \varepsilon^{\mu\nu\rho\lambda}F_{\mu\nu}^a F_{\rho\lambda}^a,
\end{align}
with $\varepsilon^{\mu\nu\rho\lambda}=\epsilon^{\mu\nu\rho\lambda}/\sqrt{-g}$
being the Levi-Civit\`{a} tensor
and $\epsilon^{0123}=1$.

A flat Friedmann-Lemaître-Robertson-Walker(FLRW) universe is compatible with the SU(2) gauge field configuration
\begin{align}
    A_0^a=0,\quad A_i^a=a(t)\psi(t)\delta_i^a.
\end{align}
In contrast,
it is not so straightforward to see whether a {\em curved} universe
is also compatible with the vector fields having nonvanishing spatial components.
However, actually it has long been known that there is a homogeneous and isotropic
SU(2) gauge field configuration for a curved
universe (see, e.g., Refs.~\cite{Hosotani:1984wj,Galtsov:1991un,Galtsov:2008wkj}).
Following these previous works, here we derive the basic equations
governing the axion and SU(2) gauge field dynamics in a spatially curved universe.

The metric for a spatially curved FLRW universe is given by
\begin{align}
  \mathrm{d} s^{2}=-N^{2}(t) \mathrm{d} t^{2}+a^{2}(t)\left[\mathrm{d} r^{2}+S^{2}(r)\left(\mathrm{d} \theta^{2}+\sin ^{2} \theta \mathrm{d} \varphi^{2}\right)\right],
\end{align}
where
\begin{align}
  S(r)=\frac{\sin (\sqrt{\mathcal{K}} r)}{\sqrt{\mathcal{K}}},
  \quad
  \frac{\sinh (\sqrt{-\mathcal{K}} r)}{\sqrt{-\mathcal{K}}}
\end{align}
for the closed ($\mathcal{K}>0$) and open ($\mathcal{K}<0$) cases, respectively.
In our notation, ${\cal K}$ has the dimension of (length)$^{-2}$.
We may always set $N=1$.

The gauge field configuration compatible with a homogeneous, isotropic, and
spatially curved FLRW universe is given as follows.
First, we fix the gauge freedom so that the time component vanishes.
Then, following~\cite{Galtsov:1991un}, we consider the ansatz
\begin{align}
  & g_A A_{0}^{a}=0,\label{ansatz A0}
  \\
  & g_A A_{1}^{a}=a\psi L_{1}^{a},\label{ansatz A1}
  \\
  & g_A A_{2}^{a}=a\psi S L_{2}^{a}-\left(1-\sqrt{1-\mathcal{K} S^{2}}\right) L_{3}^{a},\label{ansatz A2}
  \\
  & g_A A_{3}^{a}=\left[\left(1-\sqrt{1-\mathcal{K} S^{2}}\right) L_{2}^{a}+a\psi S L_{3}^{a}\right] \sin \theta ,\label{ansatz A3}
\end{align}
where $\psi=\psi(t)$ is a function of $t$ only and $L_{i}^{a}$ is defined as
\begin{align}
  & L_{1}^{a}=(\sin \theta \cos \varphi, \sin \theta \sin \varphi, \cos \theta),
  \\
  & L_{2}^{a}=(\cos \theta \cos \varphi, \cos \theta \sin \varphi,-\sin \theta),
  \\
  & L_{3}^{a}=(-\sin \varphi, \cos \varphi, 0).
\end{align}
(See also~\cite{Gorji:2019ttx} for a detailed derivation.)
For this gauge field configuration, we see that
the following terms in the Lagrangian
depend only on $t$,
\begin{align}
  &F_{\mu \nu}^{a} F_{a}^{\mu \nu} = -\frac{6}{g_A^2N^2}
  \left[\left(\dot{\psi} +\frac{\dot{a}}{a} \psi\right)^2
  -N^2 \left(\psi ^2 -\frac{\mathcal{K}}{a^2}\right)^2\right],
  \label{eq:FF01}
  \\
  &\widetilde{F}_{\mu \nu}^{a} F_{a}^{\mu \nu} = \frac{12}{g_A^2N}
  \left(\dot{\psi} +\frac{\dot{a}}{a} \psi\right)
  \left(\psi ^2 -\frac{\mathcal{K}}{a^2}\right),
  \label{eq:dFF01}
\end{align}
showing that the above ansatz is indeed consistent with
a homogeneous, isotropic, and spatially curved FLRW universe.
Here a dot denotes differentiation with respect to $t$.
Note that ${\cal K}$ appears directly in Eqs.~\eqref{eq:FF01}
and~\eqref{eq:dFF01}, which implies that the axion-SU(2) dynamics
depends nontrivially on the spatial curvature.
In other words, the effective potential for $\psi$ depends
explicitly on ${\cal K}$.
The previous spatially flat result can be reproduced
by rescaling $\psi$ as $\psi=g_A\psi_{\rm previous}$ and setting ${\cal K}=0$.
In the spatially flat case,
the rescaling of $\psi$ allows us to consider the $g_A\to 0$ limit.
However, in the case of ${\cal K}\neq 0$,
one cannot take such a smooth $g_A\to 0$ limit.
In this sense, the fact that the vector field is the SU(2) gauge field is crucial
for the above ansatz.

From the Lagrangian~\eqref{eq:Lag01} and our ansatz for the gauge field,
we obtain the field equations as
\begin{align}
  &\ddot{\chi} +3H\dot{\chi} -\frac{\mu ^4}{f} \sin \frac{\chi}{f}
  =-\frac{3\lambda}{f g_A^2} \left(\dot{\psi} +H \psi\right)
  \left(\psi ^2 -\frac{\mathcal{K}}{a^2}\right),\label{eq:axion eom}
  \\
  &\ddot{\psi}+3 H \dot{\psi}
  +\left(\dot{H}+2 H^2\right)\psi
  =\left(\frac{\lambda}{f}\dot{\chi} -2\psi\right)
  \left(\psi^2 -\frac{\mathcal{K}}{a^2}\right), \label{eq:gauge eom}
\end{align}
where $H:=\dot{a}/a$ is the Hubble parameter and we set $N=1$.
The Einstein equations read
\begin{align}
  3\mpl^2\left(H^2 +\frac{\mathcal{K}}{a^2}\right)&=
  \rho_{\chi} +\rho_{\psi},\label{eq:Friedmann eq.}
  \\
  -2\mpl^2 \left(\dot{H}-\frac{\mathcal{K}}{a^2} \right) &=
  \rho_{\chi} +\rho_{\psi} +p_{\chi} +p_{\psi},\label{eq:accel eom}
\end{align}
where
\begin{align}
  \rho_{\chi} &= \frac{\dot{\chi}^2}{2} +\mu^4\left(1+\cos\frac{\chi}{f}\right),
  \\
  p_{\chi} &= \frac{\dot{\chi}^2}{2} -\mu^4\left(1+\cos\frac{\chi}{f}\right),
  \\
  \rho_{\psi} &= \frac{3}{2g_A^2}\left[\left(\dot{\psi}
  +H \psi\right)^2 +\left(\psi^2 -\frac{\mathcal{K}}{a^2}\right)^2\right],
  \\
  p_{\psi} &=\frac{\rho_{\psi}}{3}.
\end{align}
We will solve these equations numerically
to see the nontrivial dynamics brought by the spatial curvature.

\subsection{Numerical results}\label{sec:Numerical Results: Axion-SU(2)}

For a given set of parameters we impose the initial conditions
$(\chi_0,\psi_0,\dot\chi_0,\dot\psi_0,H_0)$ at $a=a_0=1$ so that
they satisfy the Friedmann equation~\eqref{eq:Friedmann eq.}.
We then solve numerically the dynamical equations~\eqref{eq:axion eom},~\eqref{eq:gauge eom},
and~\eqref{eq:accel eom}. We confirm that the Friedmann equation is satisfied at each time step.
The parameters used in our numerical calculations
(except for the ones in which $g_A$ is varied)
are listed in \tabref{tab:parameters}.
(In the actual numerical calculations we adopt the units $\mpl=1$.)
Here we introduce for convenience the scale of the (would-be) inflationary Hubble parameter,
$\hinf$, defined as $\hinf:=\mu^2/\mpl$.
We set $\hinf$ to be the grand unified theory scale.

\begin{table}[htb]
  \centering
    \caption{Parameters}
  \label{tab:parameters}
  \begin{tabular}{|c|c|c|c||c|} \hline
    $f$ & $\mu$ & $\lambda$ & $g_A$ & $\hinf$ \\ \hline
    $10^{-2}\mpl$ & $10^{-3}\mpl$ & $2\times 10^{2}$ & $2\times 10^{-6}$ &$10^{-6}\mpl$\\ \hline
  \end{tabular}
\end{table}

We are interested in particular in the case of a closed universe.
To clarify the role of the gauge field in the inflationary dynamics with ${\cal K}>0$,
let us first consider the case without the gauge field. In this case,
the curvature term acts as effective negative energy density.
Suppose that at some initial moment the curvature term
is as large as $\mpl^2{\cal K}/a^2\sim V\sim \mu^4$.
Such a universe would typically collapse within one $e$-fold.
If the kinetic energy of the axion field is sufficiently large
(i.e., $\mpl^2H^2\sim \dot\chi^2\gg \mpl^2{\cal K}/a^2\sim V$),
the universe could avoid to collapse, but the axion would then roll down to the
potential minimum very rapidly, preventing prolonged inflation.

Let us next switch on the gauge field and investigate
the axion-SU(2) dynamics in a {\em flat} universe~\cite{Adshead:2012kp}.
It is assumed that $\dot H$, $\ddot \chi$, and $\ddot \psi$ can be ignored
in Eqs.~\eqref{eq:axion eom} and~\eqref{eq:gauge eom} (with ${\cal K}=0$).
We can then rearrange
these equations so that $\dot\chi$ and $\dot \psi$ are expressed
in terms of $\chi$ and $\psi$, and find
\begin{align}
    3H\dot\psi\simeq -V_{{\rm eff},\psi},
\end{align}
with
\begin{align}
    V_{{\rm eff},\psi}=3H^2\psi-\frac{\ga^2\mu^4H\sin(\chi/f)}{\lambda \psi^2},
\end{align}
where we assumed that
\begin{align}
 f\ga H\ll \lambda\psi^2.\label{condition-psi}
\end{align}
Although this effective potential is obtained by making some approximations
and assumptions, it is useful for understanding roughly the evolution of the system.
The minimum of the effective potential is given by
\begin{align}
  \psim = \left(\frac{\ga^2\mu^{4}}{3\lambda H} \sin \frac{\chi}{f}\right)^{1/3}.
\end{align}
One usually assumes that $\psi$ takes this value from the beginning,
$\psi\simeq \psi_{\rm min}$,
and study the dynamics of the axion $\chi$.
However, the actual dynamics of the axion depends on the initial condition for $\psi$.
If $\psi\gtrsim \psi_{\rm min}$ initially,
it rolls down to the minimum of the effective potential within a few $e$-folds,
leading to successful $\psi$-assisted inflation as
proposed originally in~\cite{Adshead:2012kp}.
If $\psi$ is displaced from $\psi_{\rm min}$ to the left
to some extent initially, then the axion field quickly falls down
into the minimum of its potential without sufficient inflation.
The reason is as follows. For sufficiently small $\psi$,
the right-hand side of Eq.~\eqref{eq:axion eom} (with ${\cal K}=0$),
i.e., the interaction term between the axion and the gauge field,
is much smaller than the bare potential of $\chi$.
Thus, $\chi$ quickly rolls down to the minimum before
$\psi$ becomes sufficiently large to assist slow-roll inflation.
This occurs when the condition~\eqref{condition-psi} is violated.

\begin{figure}[htb]
    \begin{center}
        \includegraphics[keepaspectratio=true,height=45mm]{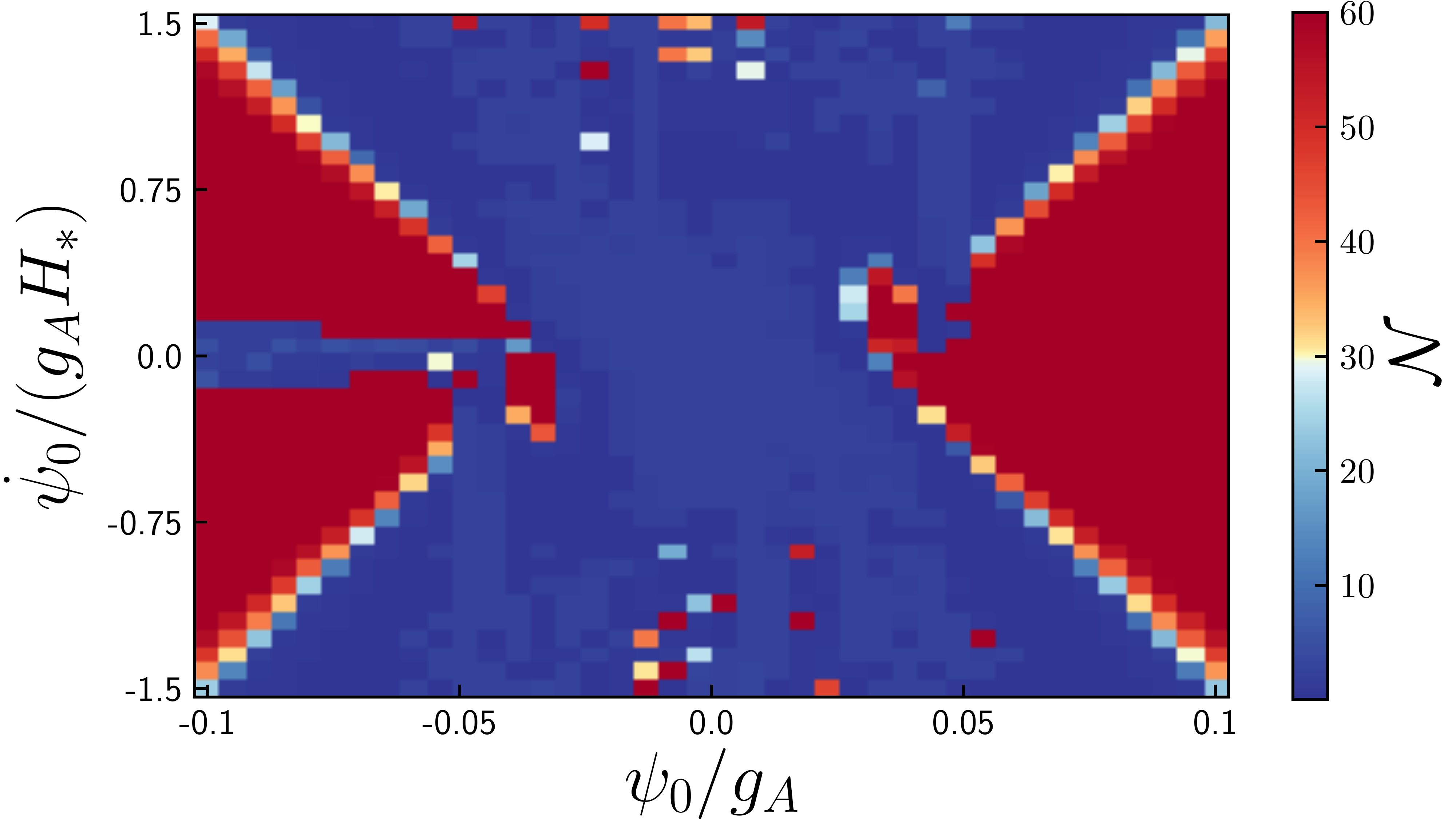}
    \end{center}
      \caption{Initial condition dependence of the $e$-folding number
      in the flat model.
      The parameters are given in~\tabref{tab:parameters}.
      }
      \label{fig:pp-dp_flat.pdf}
  \end{figure}

Figure~\ref{fig:pp-dp_flat.pdf} shows the $e$-folding number,
\begin{align}
    {\cal N}:=\int_{t_{\rm i}}^{t_{\rm f}}H\D t,
\end{align}
calculated for different initial conditions for the gauge field.
Here, $t_{\rm i}$ is the initial time and
$t_{\rm f}$ is defined as the time at which
the energy density of $\chi$ becomes as small as
$\rho_{\chi}=10^{-3}\mpl^2\hinf^2$.
If this does not occur and inflation lasts for 60 $e$-folds, we
stop the calculation and just set ${\cal N}=60$.
The initial condition for the axion is fixed as
\begin{align}
  \left[
  \begin{array}{c}
    \chi_0 \\
    \dot{\chi}_0
  \end{array}\right]=
  \left[
  \begin{array}{c}
    \pi f \times 10^{-2} \\
    0
  \end{array}
  \right].
  \label{ic_axion}
\end{align}
For our choice of the parameters, we have $\psim /\ga\simeq 0.07$,
and we can see from \figref{fig:pp-dp_flat.pdf} the
aforementioned behavior, though the dynamics depends also on
the initial value of $\dot\psi$.


We now move to the discussion on the case with ${\cal K}> 0$.
The evolution of the system
would be more complicated due to the various ${\cal K}$-dependent terms in
the basic equations.
Now positive spatial curvature does not simply act as effective negative energy density,
and hence large positive ${\cal K}$ does not necessarily make the universe collapse.
Spatial curvature also modifies the shapes of the effective potentials
for $\chi$ and $\psi$, as seen from Eqs.~\eqref{eq:axion eom}
and~\eqref{eq:gauge eom}, and thus affects directly the evolution of
these fields.

Let us estimate the effect of spatial curvature in Eq.~\eqref{eq:Friedmann eq.}.
Assuming that the curvature is large and
the kinetic energy of the fields is small enough,
Eq.~\eqref{eq:Friedmann eq.} reduces to
\begin{align}
  3\mpl^2 H^2 &\simeq \frac{3}{2\ga^2}\frac{{\cal K}^2}{a^4}
  -3\mpl^2\frac{{\cal K}}{a^2}+V,\label{FridKKV}
\end{align}
and $V\simeq 2\mu^4=2\mpl^2H_*^2$.
This can be recast into a dimensionless form as
\begin{align}
    1\simeq \frac{1}{2}\left(\frac{H}{g_A\mpl}\right)^2
    \left(\frac{{\cal K}}{a^2H^2}\right)^2
    -\frac{{\cal K}}{a^2H^2}+\frac{2}{3}\left(
    \frac{H_*}{H}\right)^2.
\end{align}
We consider the case where $H\sim H_*$ and
${\cal K}/a^2H_*^2\sim 1$.
It can then be seen that if $g_A\sim H_*/\mpl$,
the first term in the right-hand side can cancel the
second term, supporting the universe against collapse.
For larger $g_A$, the first term becomes less important.
Note that $g_A$ cannot be too small in order
for the Friedmann equation to be satisfied
in the universe with large spatial curvature.

\begin{figure}[htb]
    \begin{center}
        \includegraphics[keepaspectratio=true,height=45mm]{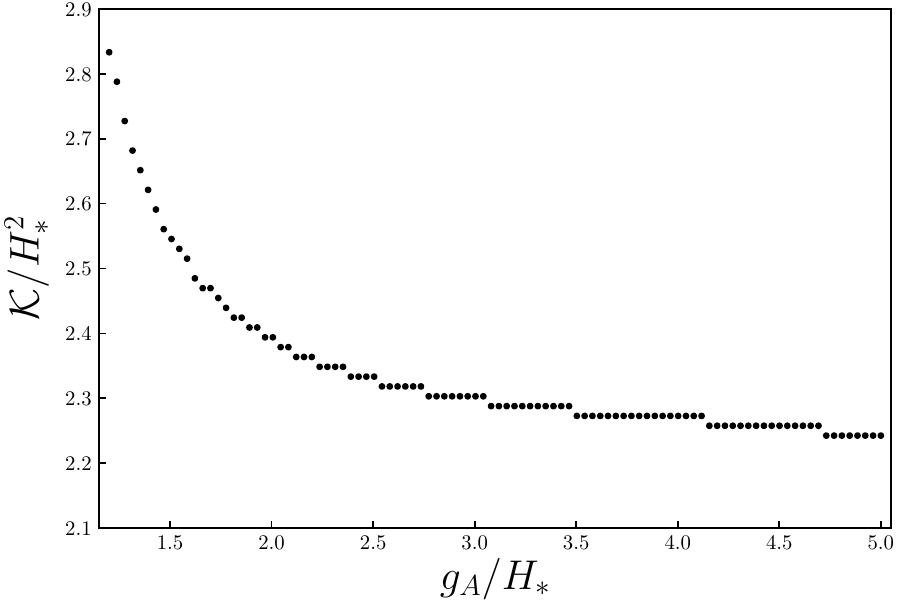}
    \end{center}
      \caption{
      The critical values of ${\cal K}$ are shown as black dots, above which
      the universe immediately collapses.
      }
      \label{fig:plot_collapse.pdf}
  \end{figure}

The above argument can be verified by the numerical result
shown in \figref{fig:plot_collapse.pdf}.
We run the numerical code for different values of ${\cal K}$ and $g_A$.
The initial conditions for the axion and the Hubble parameter
are given respectively by Eq.~\eqref{ic_axion}
and $H_0=\hinf$, while
the initial value of the gauge field is given by $\psi_0 =0.05\times \ga$.
The initial velocity of the gauge field, $\dot\psi_0$,
is then determined from the Friedmann equation~\eqref{eq:Friedmann eq.},
which is different for different $(g_A,{\cal K})$.
For ${\cal K}$ above a certain value the universe immediately collapses.
In \figref{fig:plot_collapse.pdf} this critical value of ${\cal K}$ is shown
as a function of $g_A$.
We see that the universe with smaller $g_A$
is more stable against the inclusion of
positive spatial curvature.

\begin{figure}[tb]
    \begin{center}
        \includegraphics[keepaspectratio=true,height=45mm]{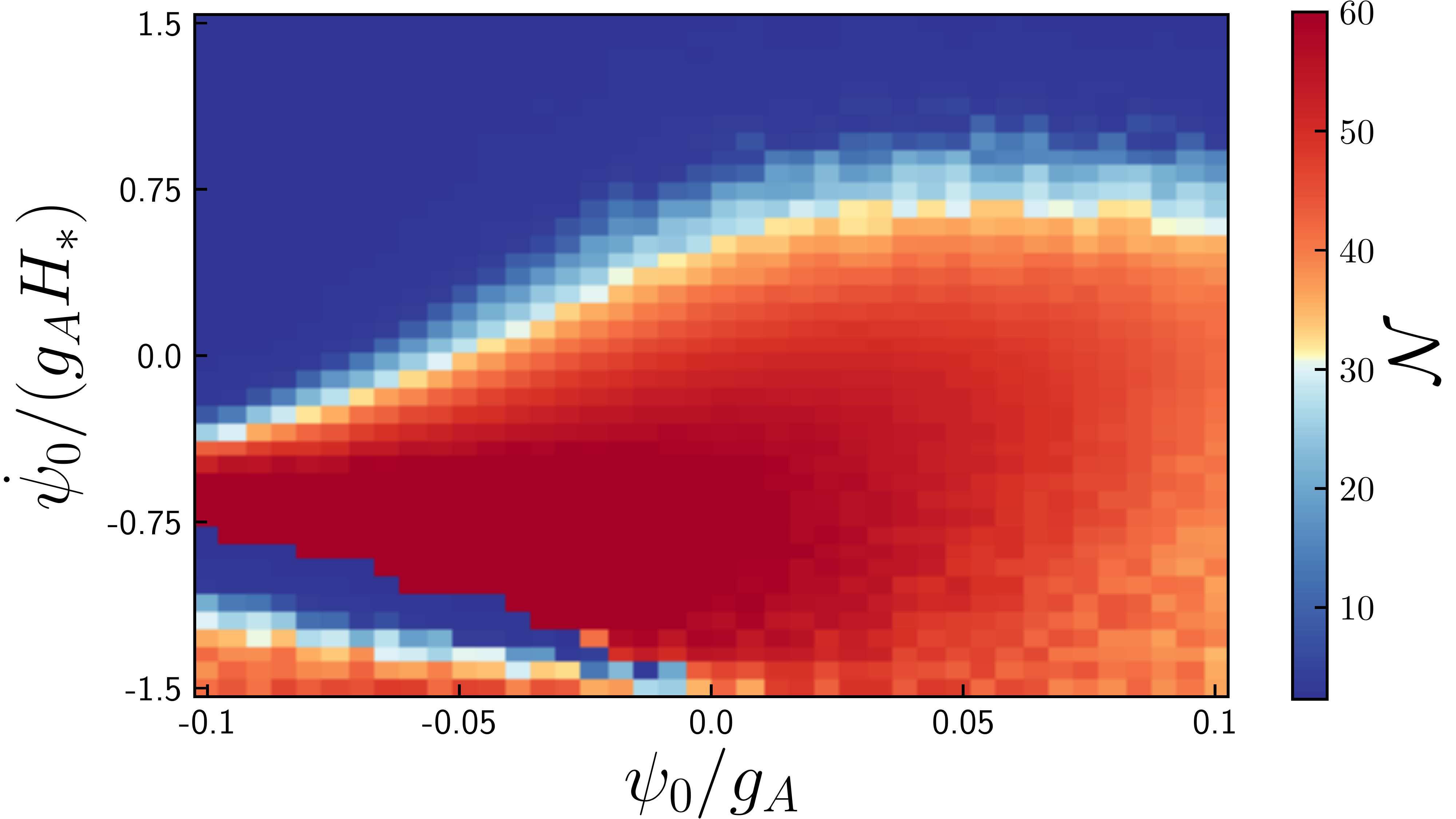}
    \end{center}
      \caption{Initial condition dependence of the $e$-folding number
      in the closed model with $\mathcal{K}=0.5\hinf ^2$.
      The initial condition for the axion field
      is given by Eq.~\eqref{ic_axion}.
  	}
      \label{fig:ic plot dp-pp}
  \end{figure}
\begin{figure}[tb]
    \begin{center}
        \includegraphics[keepaspectratio=true,height=45mm]{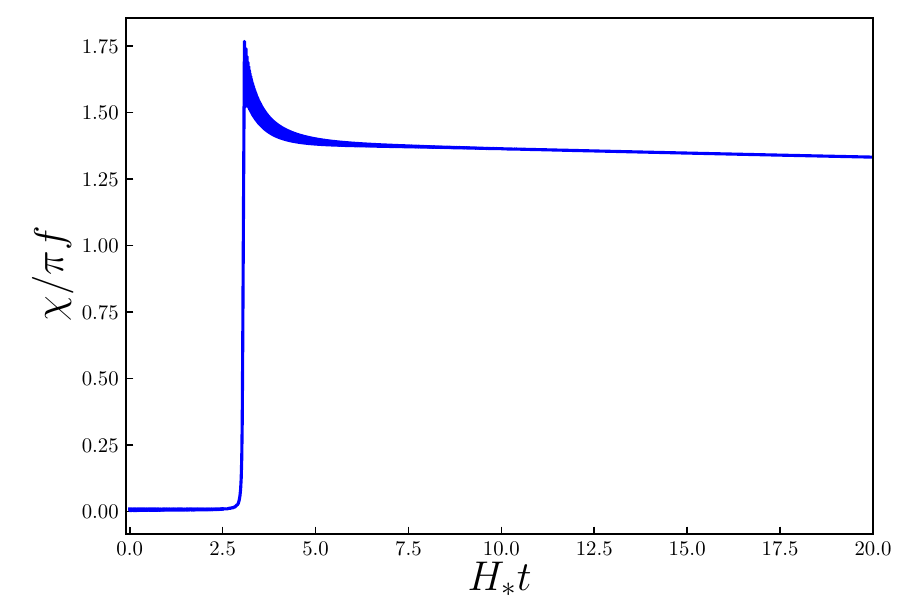}
        \includegraphics[keepaspectratio=true,height=45mm]{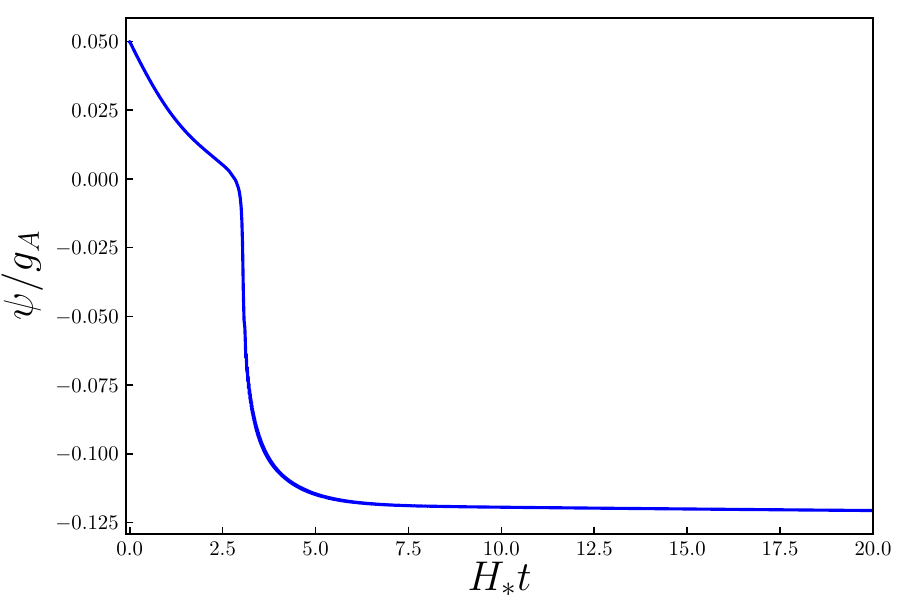}
    \end{center}
      \caption{The evolution of the axion and gauge fields
      with the curvature $\mathcal{K}=0.5\hinf^2$
      for the initial condition given by
      Eq.~\eqref{ic_axion},
      $\psi_0=10^{-7}\mpl$,
      and $\dot{\psi}_0=-10^{-12}\mpl^2$
      (for which $H_0=0.555 H_*$).}
      \label{fig:K=12chi.pdf}
  \end{figure}

The initial condition dependence of
the $e$-folding number is also altered by the presence of spatial curvature,
as can be seen by comparing \figref{fig:pp-dp_flat.pdf}
with \figref{fig:ic plot dp-pp}.
We find that, as long as the initial velocity of $\psi$ is small,
we have a sufficient duration of the inflationary phase
for a wide range of initial conditions $\psi_0$.
Note here that
the initial conditions plotted in \figref{fig:ic plot dp-pp} all satisfy
$\psi_0^2\ll{\cal K}/H_*^2$. The typical dynamics of
the axion and gauge fields in this case is shown in \figref{fig:K=12chi.pdf},
which is different from the conventional one
in the early stage, but leads in the end to prolonged inflation.
In the early stage,
the interaction term in the right-hand side of Eq.~\eqref{eq:axion eom}
is maintained even for tiny $\psi_0$
for ${\cal O}(1)$ $e$-folds,
because we have $\psi^2- {\cal K}/a^2\simeq -{\cal K}/a^2$.
Then, both fields change their values rapidly at around
$H_*t\simeq 3$. This corresponds to the moment at which
the right-hand sides of Eqs.~\eqref{eq:axion eom} and~\eqref{eq:gauge eom}
vanishes, $\psi^2\simeq {\cal K}/a^2$.
After that, $\psi$-assisted slow-roll inflation occurs
at the different side of the potential from the one
where the axion field is initially placed.
We thus have the unconventional pre-inflationary dynamics
leading to successful inflation.


So far we have considered the case where the axion is the inflaton field.
Let us briefly comment on the case where there is another scalar field that
plays the role of the inflaton and
the axion is a spectator field~\cite{Dimastrogiovanni_2017}.
Suppose in this spectator axion-SU(2) model that the inflaton's potential energy
is given by $V_{\rm inf}\sim 2\alpha \mu^4$ ($\alpha > 1$).
This amounts to rescaling $H_*$ as $H_*\to (1+\alpha)^{1/2}H_*$
in Eq.~\eqref{FridKKV}. Therefore, for larger inflaton's contribution
(i.e., for larger $\alpha$), $g_A/H_*$ becomes effectively smaller,
rendering the universe more stable against the inclusion of
positive spatial curvature.

\section{Multiple vector fields without the non-abelian-specific structure}\label{sec:proca}

Let us move to the case of multiple generalized Proca fields~\cite{Jimenez:2016upj,Emami_2017}.
In this case, the vector fields do not have the non-abelian-specific structure
and hence we cannot use the same ansatz in a spatially curved universe
as in the case of the SU(2) gauge field.
In this section, we therefore start from reconsidering the metric and
vector field ansatz.

The Lagrangian for the model considered in Ref.~\cite{Emami_2017}
contains three vector fields $A_\mu^a$ ($a=1,2,3$) and
is given by
\begin{align}
  {\cal L}=\mathcal{L}_{2}+\mathcal{L}_{4},
\end{align}
with
\begin{align}
  \mathcal{L}_{2}&=G_{2}\left(X, Y, Z, W_{1}, W_{2}, W_{3}\right),
  \\
  \mathcal{L}_{4}&=G_{4}(X) R+G_{4, X} \left[\nabla_{\mu} A^{\mu}_a \nabla^{\nu} A_{\nu}^a-\nabla_{\mu} A_{\nu}^a \nabla^{\nu} A^{\mu}_a\right],
\end{align}
where
\begin{align}
  &X := -\frac{1}{2} A_{\mu}^{a} A^{\mu}_a,
  \quad
  Y :=-\frac{1}{4} F_{\mu\nu}^{a} F^{\mu \nu}_a,
  \notag \\
  &Z :=-\frac{1}{4} F_{\mu \nu}^{a} \widetilde{F}^{\mu\nu}_a,
  \quad
  W_{1} := A_{\mu}^{a} A^{\nu}_{a} F^{\mu\rho}_b F_{\nu\rho}^b,
  \notag \\
  &W_{2} := A_{\mu}^{a} A^{\nu}_{b} F^{\mu\rho}_a F_{\nu\rho}^b,
  \quad
  W_{3} := A_{\mu}^{a} A^{\nu}_{b} F^{\mu\rho}_b F_{\nu\rho}^a,
\end{align}
and note that in the present case
\begin{align}
    F_{\mu \nu}^{a}:=\partial_{\mu} A_{\nu}^{a}-\partial_{\nu} A_{\mu}^{a}.
\end{align}

To explore the configuration of the vector fields,
we first note that the metric of a closed universe can be written
using the left-invariant 1-forms as
\begin{align}
  \mathrm{d} s^{2}=-N^{2}(t) \mathrm{d} t^{2}+\frac{a^2(t)}{4}
  \delta_{ab} \omega^{a} \omega^{b},\label{met:vec3}
\end{align}
where
\begin{align}
  \omega^{1}&=-\sin \left(\sqrt{\mathcal{K}} x^{3}\right) \mathrm{d} x^{1}+\sin \left(\sqrt{\mathcal{K}} x^{1}\right) \cos \left(\sqrt{\mathcal{K}} x^{3}\right) \mathrm{d} x^{2}, \\
  \omega^{2}&=\cos \left(\sqrt{\mathcal{K}} x^{3}\right) \mathrm{d} x^{1}+\sin \left(\sqrt{\mathcal{K}} x^{1}\right) \sin \left(\sqrt{\mathcal{K}} x^{3}\right) \mathrm{d} x^{2}, \\
  \omega^{3}&=\cos \left(\sqrt{\mathcal{K}} x^{1}\right) \mathrm{d} x^{2}+\mathrm{d} x^{3}.
\end{align}
Notice that this expression can be used only for a closed universe.
We then assume that the vector fields take the form\footnote{In Appendix~\ref{app:1},
we show that the same ansatz can also be used for the SU(2) gauge field, reproducing
the same result as in Sec.~\ref{sec:su2curve} up to field redefinition.}
\begin{align}
  A_{0}^{a}=0, \quad A_{i}^{a} \mathrm{d} x^{i}=\frac{a}{2} \psi(t)\omega^{a},
  \label{eq:ansatz3vec}
\end{align}
where we set $A_0^a=0$ because it is indeed a trivial solution.
Under the above ansatz we have
\begin{align}
  X &= -\frac{3}{2} \psi^{2},\label{eq:Xt}
  \\
  Y &= \frac{3}{2}\left[\frac{1}{N^{2}}\left(\dot{\psi}+\frac{\dot{a}}{a} \psi\right)^{2}-\frac{4 \mathcal{K}}{a^{2}} \psi^{2}\right],\label{eq:Yt}
  \\
  Z &= \frac{6}{N} \psi \left(\dot{\psi}+\frac{\dot{a}}{a} \psi\right) \frac{\sqrt{\mathcal{K}}}{a},
  \\
  W_{1} &= -3\psi ^2\left[\frac{1}{N^{2}}\left(\dot{\psi}+\frac{\dot{a}}{a} \psi\right)^2 -\frac{8 \mathcal{K}}{a^2}\psi ^2\right],
  \\
  W_{2} &= -\frac{9}{N^{2}} \psi ^2\left(\dot{\psi}+\frac{\dot{a}}{a} \psi\right)^{2},
  \\
  W_{3} &= -3\psi ^2\left[\frac{1}{N^{2}}\left(\dot{\psi}+\frac{\dot{a}}{a} \psi\right)^2
  +\frac{8 \mathcal{K}}{a^2}\psi ^2\right],
\end{align}
and
\begin{align}
    \nabla_{\mu} A^{\mu}_a \nabla^{\nu} A_{\nu}^a-\nabla_{\mu} A_{\nu}^a \nabla^{\nu} A^{\mu}_a
    =
    -\frac{6}{N^2}\frac{\dot{a}}{a} \dot{\psi}\psi +\frac{6\mathcal{K}}{a^2}\psi^2.
\end{align}
We see that each piece in the
Lagrangian depends only on $t$, and therefore the
above assumed configuration of the vector fields is consistent
with a closed universe.
In contrast to the spatially flat case~\cite{Emami_2017},
$Z$ is no longer vanishing in a positively curved universe.

Let us present two applications of the our result.
The first example is the vector inflation model
proposed in Ref.~\cite{Golovnev:2008cf}, whose Lagrangian is given by
\begin{align}
    {\cal L}=\left(\frac{\mpl^2}{2}-\xi X\right)R+Y+m^2X,
    \label{Lag:vector-inf08}
\end{align}
where $\xi$ and $m$ are constants, and below we
will set $\xi=1/6$. Strictly speaking, this does not
belong to the multiple generalized Proca theory,
as the Lagrangian~\eqref{Lag:vector-inf08} is obtained by
omitting $G_{4,X}$ by hand~\cite{Emami_2017}.
Nevertheless, one can use the ansatz~\eqref{eq:ansatz3vec}
to analyze the dynamic of the vector inflation model of~\cite{Golovnev:2008cf}
with ${\cal K}>0$, since, as seen from Eqs.~\eqref{eq:Xt} and~\eqref{eq:Yt},
both $X$ and $Y$ are dependent only on $t$.
Our field equations are given by
\begin{align}
    &\ddot{\psi} +3H\dot{\psi} +\left(m^2 +\frac{3\mathcal{K}}{a^2}\right)\psi=0,
    \label{vecinf01}
    \\
    &3\mpl^2\left(H^2+\frac{\mathcal{K}}{a^2}\right) =\frac{3}{2} \left(\dot{\psi}^2 +m^2 \psi^2\right) +\frac{9\mathcal{K}}{2a^2}\psi^2,
    \\
    &2\mpl^2 \left(\dot{H}- \frac{\mathcal{K}}{a^2}\right) =- 3\dot{\psi}^2 -\frac{3\mathcal{K}}{a^2} \psi^2.\label{vecinf03}
\end{align}
Open and closed FLRW models in the theory~\eqref{Lag:vector-inf08}
have been studied earlier in Ref.~\cite{Chiba:2008eh}.
Our result disagrees with that of Ref.~\cite{Chiba:2008eh};
it seems that the consistent ansatz for the vector fields
has not been employed in~\cite{Chiba:2008eh}.
The sign of the vector-field-induced curvature terms in Eqs.~\eqref{vecinf01}--\eqref{vecinf03}
is opposite compared to the result of~\cite{Chiba:2008eh},
and accordingly
the dynamics should be qualitatively different
as compared to~\cite{Chiba:2008eh}.

The second example is the model with a triplet of vector fields coupled to a scalar field.
Obviously, the ansatz~\eqref{eq:ansatz3vec} can be used in such a case as well.
For example,
the model studied in Ref.~\cite{Gorji_2020} is described by the Lagrangian
\begin{align}
  \mathcal{L} = \frac{\mpl^2}{2}R-\frac{1}{2}(\partial\phi)^2-V(\phi)+
  f^2(\phi)\left(Y +\theta Z \right),
\end{align}
where $\theta$ is a constant.
In this case, the vector fields have the
U(1)$\times$U(1)$\times$U(1) symmetry.
(One may thus use this symmetry to eliminate $A_0^a$.)
Let us investigate the cosmological background dynamics of this model
in more detail.
The field equations are given by
\begin{align}
    &\ddot{\phi} +3H\dot{\phi} +V' -3ff'(\dot{\psi} +H\psi)^2\notag\\
    &+12ff'\psi\left[\frac{{\cal K}}{a^2}\psi
    +\theta\frac{\sqrt{\cal K}}{a} (\dot{\psi} +H\psi)\right]=0,\label{eq:f(y+z) inflaton eom}
    \\
    &\ddot{\psi} +3 H \dot{\psi} +\left(\dot H+2 H^2\right)\psi +\frac{2 f'}{f} \dot{\phi}\left(\dot{\psi} +H \psi\right)\notag\\
    &+4\psi\left(\frac{\mathcal{K}}{a^2} -\theta\frac{\sqrt{\mathcal{K}} }{a}\frac{f'}{f}\dot{\phi}\right)=0,
    \\
    &3\mpl^2\left(H^2+\frac{\mathcal{K}}{a^2}\right) = \frac{\dot{\phi}^2}{2} +V
    +\frac{6 \mathcal{K}}{a^2} f^2 \psi ^2 \notag\\
    & \qquad\qquad \qquad \qquad \quad
    +\frac{3}{2} f^2 \left(\dot{\psi} +H \psi\right)^2,
    \\
    &\mpl^2\left(\dot{H} -\frac{\cal K}{a^2}\right) =-\frac{\dot{\phi}^2}{2}
    -\frac{4 \mathcal{K}}{a^2}f^2 \psi^2 
    -f^2 \left(\dot{\psi} +H \psi \right)^2,
\end{align}
where a prime stands for differentiation with respect to $\phi$.
In the following,
we will consider the particular form of
the potential and the coupling function given by
\begin{align}
    V = \frac{1}{2}m^2\phi^2,
    \quad
    f=f_*\exp \left[\frac{2\mpl^{-2}}{1-I} \int^{\phi}_{\phi_*} \frac{V(\widetilde{\phi})}{V'(\widetilde{\phi})} \D \widetilde{\phi}\right],
\end{align}
where $m$, $I$, $f_*$, and $\phi_*$ are constants.
This particular case was also studied in Ref.~\cite{Gorji_2020}
and shown to admit an inflationary attractor solution
with nonvanishing vector fields~\cite{Firouzjahi:2018wlp}.

For a fixed value of ${\cal K}/a_0^2H_0^2$,
we numerically solve the field equations to see
whether the universe continues to expand toward
the inflationary attractor or stops expanding at a certain moment
and eventually collapses.
Let us first switch off the vector fields by setting $f_*=0$
and consider the case where ${\cal K}/a_0^2H_0^2=7.5$ at the initial
moment. We take $m=10^{-2}\mpl$.
It is easy to see that for the initial condition
$(\phi_0,\dot\phi_0)=(10\mpl,-0.1\mpl^2)$
the universe stops expanding within one $e$-fold and then collapses.
We then include the vector fields with $f(\phi_0)=0.1$, $I=0.1$, and $\theta=0$.
For the initial condition
$(\phi_0,\dot\phi_0,\psi_0,\dot\psi_0)=(10\mpl,-0.1\mpl^2,\mpl,0)$,
we find that the universe expands toward the inflationary attractor
rather than collapses even in the presence of such
a large spatial curvature at the initial moment.
This implies that the triplet of vector fields
can save the universe from collapsing.

\begin{figure}[htb]
    \begin{center}
            \includegraphics[keepaspectratio=true,height=60mm]{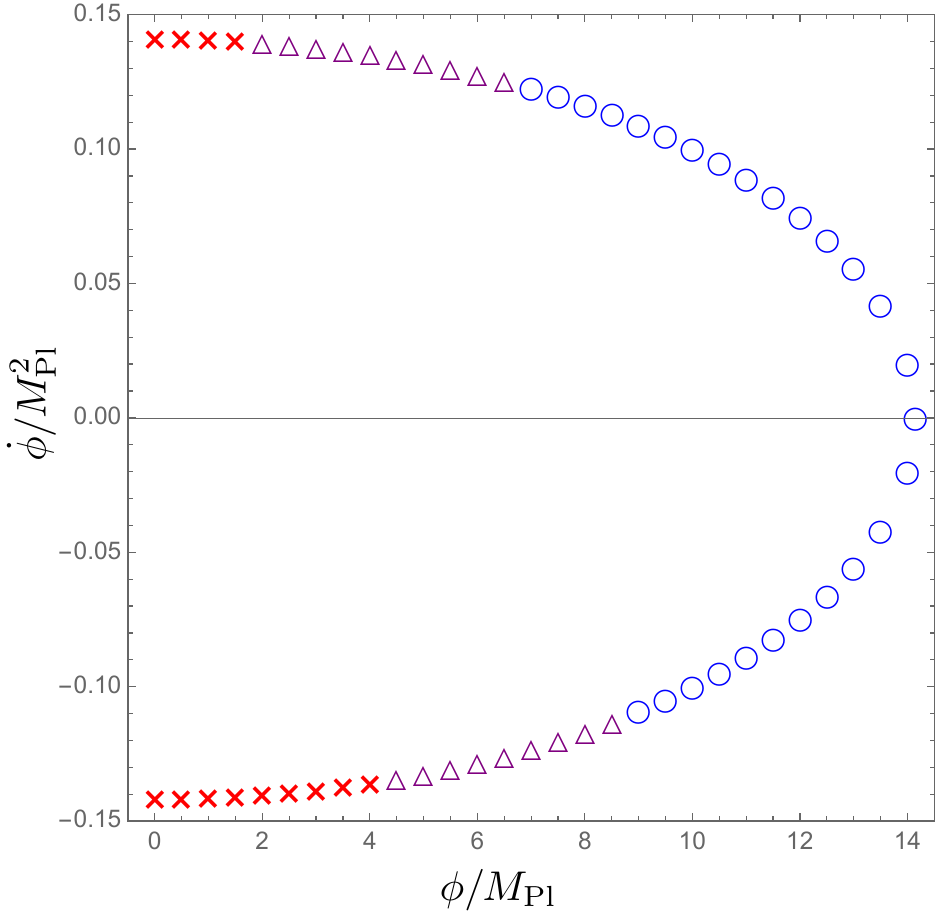}
    \end{center}
      \caption{The circles represent the initial conditions that result in inflation both with and without the vector fields, while
      for the initial conditions represented by
      the crosses the universe immediately collapses in any case.
      For the initial conditions represented by the triangles,
      the universe without the vector fields collapses,
      but with the help of the vector fields it avoids to do so,
      leading to successful inflation.
      No initial conditions are found for which
      the vector fields hinder inflation.
      Due to the symmetry only the cases with $\phi_0>0$ are shown.
  	}
      \label{fig:f(Y+Z)_point.pdf}
  \end{figure}

Figure~\ref{fig:f(Y+Z)_point.pdf} shows the numerical results
for different initial conditions satisfying
$(1/2)\dot\phi_0^2+V(\phi_0)=10^{-2}\mpl^4$,
with $(\psi_0,\dot\psi_0)=(\mpl,0)$.
The parameters are the same as above:
${\cal K}/a_0^2H_0^2=7.5$, $m=10^{-2}\mpl$,
$I=0.1$, and $\theta=0$. For each calculation we introduce
$f_*$ and $\phi_*$ so that $f(\phi_0)=0.1$,
We see that the fraction of initial conditions
leading to successful inflation indeed increases with the
help of the vector fields, though the effect is not so
significant.

\begin{figure}[htb]
    \begin{center}
            \includegraphics[keepaspectratio=true,height=60mm]{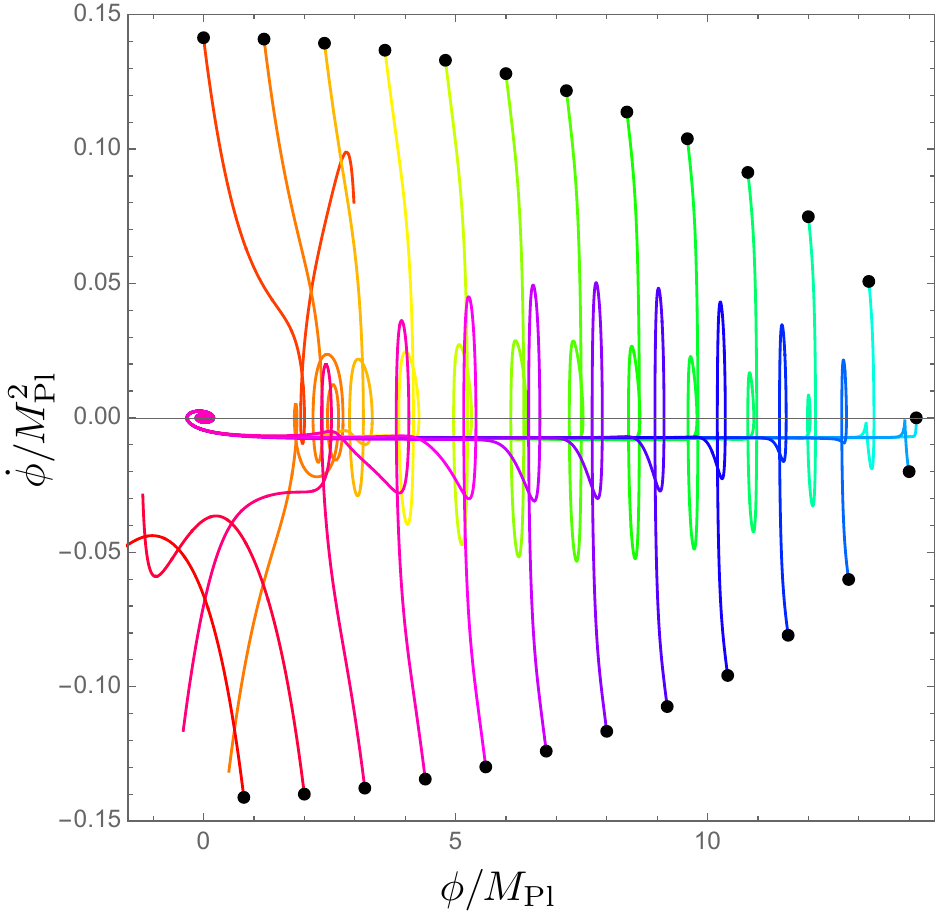}
    \end{center}
      \caption{Trajectories of the inflaton in the phase space. Black points represent the initial conditions.
  	}
      \label{fig:f(Y+Z)_trajectory.pdf}
  \end{figure}

In \figref{fig:f(Y+Z)_trajectory.pdf}, we present typical phase space trajectories
in the presence of the vector fields.
It can be seen that the dynamics of the inflaton
shows an oscillatory nature during the early stage
in which the effect of the spatial curvature is large.
This pre-inflationary dynamics is caused by
the interplay among the inflaton, the vector fields, and ${\cal K}$.
In this phase, the inflaton equation of motion~\eqref{eq:f(y+z) inflaton eom}
is given approximately by
\begin{align}
    \ddot\phi+3H\dot\phi\simeq 3ff'\left(
    \dot\psi^2-\frac{4{\cal K}}{a^2}\psi^2
    \right),
\end{align}
and
the sign of the right-hand side flips several times
because $\psi$ oscillates around zero,
yielding the oscillatory dynamics as shown in \figref{fig:f(Y+Z)_trajectory.pdf}.


\section{Conclusions and discussion}\label{sec:conclusions}

In this paper, we have studied the dynamics of
a homogeneous, isotropic, and positively curved universe
in the presence of the SU(2) gauge field or
a triplet of mutually orthogonal vector fields,
and obtained the following results.
\begin{itemize}
    \item We have obtained the consistent ansatz for
    a triplet of the generalized Proca fields
    in a closed universe, which corrects the previous analysis of
    a certain vector-field model of inflation in a curved universe.
    Our ansatz can also be used to the SU(2) gauge field
    and reproduces the known result up to field redefinition.
    \item In addition to the usual curvature terms in the cosmological
    equations, new curvature-dependent terms appear through the
    vector fields.
    In the cases of axion-SU(2) inflation~\cite{Adshead:2012kp} and
    inflation with the vector fields having U(1)$\times$U(1)$\times$U(1)
    symmetry~\cite{Gorji_2020}, we have found that the vector fields
    support the closed universe against collapse, though the
    effect is not so significant.
    \item In both models we have found nontrivial pre-inflationary dynamics
    caused by spatial curvature.
\end{itemize}

Let us discuss a possible extension of the present study.
It is easy to see that the ansatz we have
introduced in Sec.~\ref{sec:proca} can be generalized straightforwardly to
include spatial anisotropies
\begin{align}
    &\D s^2=-N^2(t)\D t^2+h_{ab}\omega^a\omega^b,
    \\
    &h_{ab}=\frac{a^2}{4}\times{\rm diag}
    (e^{-4\sigma_+},e^{2\sigma_++2\sqrt{3}\sigma_-},e^{2\sigma_+-2\sqrt{3}\sigma_-}),
    \\
    &A_i^a\D x^i=\sqrt{h_{aa}}\psi_a\omega^a
    \quad (\text{no summation over } a)
    ,
    \\
    &\psi_a=\psi(t)\times (e^{-2\beta_+},e^{\beta_++\sqrt{3}\beta_-},e^{\beta_+-\sqrt{3}\beta_-}),
\end{align}
where $\sigma_\pm=\sigma_\pm(t)$ and $\beta_\pm=\beta_\pm(t)$ characterize anisotropies.
It would therefore be
interesting to explore the Bianchi type-IX dynamics
in the presence of mutually orthogonal vector fields,
which is left for further study.

\acknowledgments
TK was partially supported by JSPS KAKENHI Grants No.~JP20H04745 and
No.~JP20K03936.

\appendix

\section{Ansatz~\eqref{eq:ansatz3vec} applied to the SU(2) gauge field}\label{app:1}

In this Appendix let us confirm that the ansatz~\eqref{eq:ansatz3vec}
can also be used for the SU(2) gauge field
and the same result as in Sec.~\ref{sec:su2curve}
is reproduced up to field redefinition (for ${\cal K}>0$).
Using Eqs.~\eqref{met:vec3} and~\eqref{eq:ansatz3vec} we find
\begin{align}
    F_{\mu\nu}^aF_a^{\mu\nu}&=-6
    \left[\frac{1}{N^2}\left(\dot\psi+\frac{\dot a}{a}\psi\right)^2
    -\psi^2\left(g_A\psi+\frac{2\sqrt{{\cal K}}}{a}\right)^2\right],
    \label{appeq:FF02}
    \\
    \notag\\
    \notag\\
    \widetilde F^a_{\mu\nu}F_a^{\mu\nu}&=-\frac{12}{N}
    \left(\dot\psi+\frac{\dot a}{a}\psi\right)\psi\left(g_A\psi
    +\frac{2\sqrt{{\cal K}}}{a}\right).
    \label{appeq:dFF02}
\end{align}
These are apparently different from Eqs.~\eqref{eq:FF01} and~\eqref{eq:dFF01}.
It is easy, however, to see that in terms of the new variable
\begin{align}
    \psi_{\rm new}:=-g_A\psi-\frac{\sqrt{{\cal K}}}{a},
\end{align}
Eqs.~\eqref{appeq:FF02} and~\eqref{appeq:dFF02}
coincide with Eqs.~\eqref{eq:FF01} and~\eqref{eq:dFF01}, respectively.

In the context of FLRW cosmologies in the presence of gauge sectors, the
same ansatz for the gauge field has been studied in Ref.~\cite{Moniz:1990hf}
and its application can be found in Ref.~\cite{VargasMoniz:2002gj}

\bibliography{refs}
\bibliographystyle{JHEP}
\end{document}